\documentclass[12pt]{article}
\usepackage{epsfig}
\usepackage{amsmath}
\usepackage{hhline}
\usepackage{amssymb}
\usepackage{times}
\usepackage{cite}
\usepackage[]{lineno}
\usepackage{hyperref} 

\newlength{\dinwidth}
\newlength{\dinmargin}
\begin{document}  
\newcommand{\pom}{{I\!\!P}}
\newcommand{\reg}{{I\!\!R}}

\newcommand{\slowpi}{\pi_{\mathit{slow}}}
\newcommand{\fiidiii}{F_2^{D(3)}}
\newcommand{\fiidiiiarg}{\fiidiii\,(\beta,\,Q^2,\,x)}
\newcommand{\n}{1.19\pm 0.06 (stat.) \pm0.07 (syst.)}
\newcommand{\nz}{1.30\pm 0.08 (stat.)^{+0.08}_{-0.14} (syst.)}
\newcommand{\fiidiiiful}{F_2^{D(4)}\,(\beta,\,Q^2,\,x,\,t)}
\newcommand{\fiipom}{\tilde F_2^D}
\newcommand{\ALPHA}{1.10\pm0.03 (stat.) \pm0.04 (syst.)}
\newcommand{\ALPHAZ}{1.15\pm0.04 (stat.)^{+0.04}_{-0.07} (syst.)}
\newcommand{\fiipomarg}{\fiipom\,(\beta,\,Q^2)}
\newcommand{\pomflux}{f_{\pom / p}}
\newcommand{\nxpom}{1.19\pm 0.06 (stat.) \pm0.07 (syst.)}
\newcommand {\gapprox}
   {\raisebox{-0.7ex}{$\stackrel {\textstyle>}{\sim}$}}
\newcommand {\lapprox}
   {\raisebox{-0.7ex}{$\stackrel {\textstyle<}{\sim}$}}
\def\gsim{\,\lower.25ex\hbox{$\scriptstyle\sim$}\kern-1.30ex%
\raise 0.55ex\hbox{$\scriptstyle >$}\,}
\def\lsim{\,\lower.25ex\hbox{$\scriptstyle\sim$}\kern-1.30ex%
\raise 0.55ex\hbox{$\scriptstyle <$}\,}
\newcommand{\pomfluxarg}{f_{\pom / p}\,(x_\pom)}
\newcommand{\dsf}{\mbox{$F_2^{D(3)}$}}
\newcommand{\dsfva}{\mbox{$F_2^{D(3)}(\beta,Q^2,x_{I\!\!P})$}}
\newcommand{\dsfvb}{\mbox{$F_2^{D(3)}(\beta,Q^2,x)$}}
\newcommand{\dsfpom}{$F_2^{I\!\!P}$}
\newcommand{\gap}{\stackrel{>}{\sim}}
\newcommand{\lap}{\stackrel{<}{\sim}}
\newcommand{\fem}{$F_2^{em}$}
\newcommand{\tsnmp}{$\tilde{\sigma}_{NC}(e^{\mp})$}
\newcommand{\tsnm}{$\tilde{\sigma}_{NC}(e^-)$}
\newcommand{\tsnp}{$\tilde{\sigma}_{NC}(e^+)$}
\newcommand{\st}{$\star$}
\newcommand{\sst}{$\star \star$}
\newcommand{\ssst}{$\star \star \star$}
\newcommand{\sssst}{$\star \star \star \star$}
\newcommand{\tw}{\theta_W}
\newcommand{\sw}{\sin{\theta_W}}
\newcommand{\cw}{\cos{\theta_W}}
\newcommand{\sww}{\sin^2{\theta_W}}
\newcommand{\cww}{\cos^2{\theta_W}}
\newcommand{\trm}{m_{\perp}}
\newcommand{\trp}{p_{\perp}}
\newcommand{\trmm}{m_{\perp}^2}
\newcommand{\trpp}{p_{\perp}^2}
\newcommand{\alp}{\alpha_s}

\newcommand{\alps}{\alpha_s}
\newcommand{\sqrts}{$\sqrt{s}$}
\newcommand{\LO}{$O(\alpha_s^0)$}
\newcommand{\Oa}{$O(\alpha_s)$}
\newcommand{\Oaa}{$O(\alpha_s^2)$}
\newcommand{\PT}{p_{\perp}}
\newcommand{\JPSI}{J/\psi}
\newcommand{\sh}{\hat{s}}
\newcommand{\uh}{\hat{u}}
\newcommand{\MP}{m_{J/\psi}}
\newcommand{\PO}{I\!\!P}
\newcommand{\xbj}{x}
\newcommand{\xpom}{x_{\PO}}
\newcommand{\ttbs}{\char'134}
\newcommand{\xpomlo}{3\times10^{-4}}  
\newcommand{\xpomup}{0.05}  
\newcommand{\dgr}{^\circ}
\newcommand{\pbarnt}{\,\mbox{{\rm pb$^{-1}$}}}
\newcommand{\gev}{\,\mbox{GeV}}
\newcommand{\WBoson}{\mbox{$W$}}
\newcommand{\fbarn}{\,\mbox{{\rm fb}}}
\newcommand{\fbarnt}{\,\mbox{{\rm fb$^{-1}$}}}
\newcommand{\dsdx}[1]{$d\sigma\!/\!d #1\,$}
\newcommand{\eV}{\mbox{e\hspace{-0.08em}V}}
%
%
\newcommand{\qsq}{\ensuremath{Q^2} }
\newcommand{\gevsq}{\ensuremath{\mathrm{GeV}^2} }
\newcommand{\et}{\ensuremath{E_t^*} }
\newcommand{\rap}{\ensuremath{\eta^*} }
\newcommand{\gp}{\ensuremath{\gamma^*}p }
\newcommand{\dsiget}{\ensuremath{{\rm d}\sigma_{ep}/{\rm d}E_t^*} }
\newcommand{\dsigrap}{\ensuremath{{\rm d}\sigma_{ep}/{\rm d}\eta^*} }

\newcommand{\dstar}{\ensuremath{D^*}}
\newcommand{\dstarp}{\ensuremath{D^{*+}}}
\newcommand{\dstarm}{\ensuremath{D^{*-}}}
\newcommand{\dstarpm}{\ensuremath{D^{*\pm}}}
\newcommand{\zDs}{\ensuremath{z(\dstar )}}
\newcommand{\Wgp}{\ensuremath{W_{\gamma p}}}
\newcommand{\ptds}{\ensuremath{p_t(\dstar )}}
\newcommand{\etads}{\ensuremath{\eta(\dstar )}}
\newcommand{\ptj}{\ensuremath{p_t(\mbox{jet})}}
\newcommand{\ptjn}[1]{\ensuremath{p_t(\mbox{jet$_{#1}$})}}
\newcommand{\etaj}{\ensuremath{\eta(\mbox{jet})}}
\newcommand{\detadsj}{\ensuremath{\eta(\dstar )\, \mbox{-}\, \etaj}}

\def\Journal#1#2#3#4{{#1} {\bf #2} (#3) #4}
\def\NCA{\em Nuovo Cimento}
\def\NIM{\em Nucl. Instrum. Methods}
\def\NIMA{{\em Nucl. Instrum. Methods} {\bf A}}
\def\NPB{{\em Nucl. Phys.}   {\bf B}}
\def\PLB{{\em Phys. Lett.}   {\bf B}}
\def\PRL{\em Phys. Rev. Lett.}
\def\PRD{{\em Phys. Rev.}    {\bf D}}
\def\ZPC{{\em Z. Phys.}      {\bf C}}
\def\EJC{{\em Eur. Phys. J.} {\bf C}}
\def\CPC{\em Comp. Phys. Commun.}

\begin{titlepage}

\noindent
\begin{flushleft}
{\tt DESY 17-043    \hfill    ISSN 0418-9833} \\
{\tt March 2017}                              \\
\end{flushleft}

\vspace{2cm}
\begin{center}
\begin{Large}

{\bf Measurement of \boldmath{$\dstar$} Production in Diffractive Deep Inelastic Scattering at HERA}

\vspace{2cm}

H1 Collaboration

\end{Large}
\end{center}

\vspace{2cm}

\begin{abstract}
Measurements of $\dstar(2010)$ meson production in diffractive deep inelastic scattering $(5<Q^{2}<100~{\rm GeV}^{2})$ are presented which are based on HERA data recorded at a centre-of-mass energy $\sqrt{s} = 319{\rm~GeV}$ with an integrated luminosity of $287$ pb$^{-1}$. The reaction $ep \rightarrow eXY$ is studied, where the system $X$, containing at least one $\dstar(2010)$ meson, is separated from a leading low-mass proton dissociative system $Y$ by a large rapidity gap. The kinematics of $\dstar$ candidates are reconstructed in the $\dstar\rightarrow K \pi\pi$ decay channel. The measured cross sections compare favourably with next-to-leading order QCD predictions, where charm quarks are produced via boson-gluon fusion. The charm quarks are then independently fragmented to the $\dstar$ mesons. The calculations rely on the collinear factorisation theorem and are based on diffractive parton densities previously obtained by H1 from fits to inclusive diffractive cross sections. The data are further used to determine the diffractive to inclusive $\dstar$ production ratio in deep inelastic scattering.
\end{abstract}

\vspace{1.5cm}

\vspace*{10mm}
\begin{center}
    Submitted to {\it Eur. Phys. J.} {\bf C}
\end{center}


\end{titlepage}

%
%
%
\begin{flushleft}

V.~Andreev$^{19}$,             
A.~Baghdasaryan$^{31}$,        
K.~Begzsuren$^{28}$,           
A.~Belousov$^{19}$,            
A.~Bolz$^{12}$,                
V.~Boudry$^{22}$,              
G.~Brandt$^{41}$,              
V.~Brisson$^{21}$,             
D.~Britzger$^{10}$,            
A.~Buniatyan$^{2}$
% ...


\bigskip{\it
 $ ^{1}$ I. Physikalisches Institut der RWTH, Aachen, Germany \\
 $ ^{2}$ School of Physics and Astronomy, University of Birmingham,
          Birmingham, UK$^{ b}$ \\
% ...
$ ^{10}$ DESY, Hamburg, Germany \\
% ...
$ ^{12}$ Physikalisches Institut, Universit\"at Heidelberg,
          Heidelberg, Germany$^{ a}$ \\
% ...
$ ^{19}$ Lebedev Physical Institute, Moscow, Russia \\
% ...
$ ^{21}$ LAL, Universit\'e Paris-Sud, CNRS/IN2P3, Orsay, France \\
 $ ^{22}$ LLR, Ecole Polytechnique, CNRS/IN2P3, Palaiseau, France \\
% ...
$ ^{28}$ Institute of Physics and Technology of the Mongolian
          Academy of Sciences, Ulaanbaatar, Mongolia \\
% ...
$ ^{31}$ Yerevan Physics Institute, Yerevan, Armenia \\
% ...
$ ^{41}$ Now at II. Physikalisches Institut, Universit\"at G\"ottingen,
          G\"ottingen, Germany \\
% ...
$ ^a$ Supported by the Bundesministerium f\"ur Bildung und Forschung, FRG,
      under contract numbers 05H09GUF, 05H09VHC, 05H09VHF,  05H16PEA \\
 $ ^b$ Supported by the UK Science and Technology Facilities Council,
      and formerly by the UK Particle Physics and
      Astronomy Research Council \\
% ...
}

\end{flushleft}
%

\newpage

\section{Introduction}

In the framework of Regge theory of soft hadronic interactions, the energy dependence of total hadron-hadron scattering cross sections is described only after taking into account a specific type of effective exchange with vacuum quantum numbers~\cite{Donnachie:1992ny}. Although it is used in various contexts, such an exchange is often referred to as a `pomeron' ($\pom$)~\cite{pomeron}. 
Pomeron exchange is a tool to describe diffractive processes, which are characterised by large gaps, devoid of activity, in the rapidity distribution of final state particles.

Diffractive processes in electron-proton\footnote{The term electron is referring to both $e^{-}$ and $e^{+}$ unless stated otherwise.} deep inelastic scattering were observed already in the very early part of the HERA experimental program~\cite{Derrick:1993xh,Ahmed:1995ns} and lead to revived interest in this class of soft peripheral hadronic interactions~\cite{Levonian:1996zi}. In reactions of the type $ep \rightarrow eXY$ they are characterised by a large gap in rapidity between the systems $X$ and $Y$. The system $X$ can be considered as resulting from  a diffractive dissociation of the virtual photon, while the system $Y$ consists of the initial state proton or its low mass hadronic excitation, scattered at a small momentum transfer squared $t$ relative to the initial state proton.

Perturbative quantum chromodynamics (pQCD) calculations are applicable in deep inelastic scattering even though the  partonic structure of the proton is {\em a priori} unknown. In order to overcome this difficulty, the collinear factorisation theorem~\cite{Collins:1989gx} is used, where the calculation of deep inelastic scattering (DIS) cross sections is described by a process-dependent partonic hard scattering part convoluted with a universal set of parton distribution functions of the proton (PDF). Collinear factorisation, therefore, opens the possibility to extract PDFs from one process and use them to predict cross sections for another process. For the PDF extraction the validity of the DGLAP evolution equations~\cite{Gribov:1972ri,Dokshitzer:1977sg,Altarelli:1977zs} is assumed.

A similar strategy can also be adapted to diffractive deep inelastic scattering (DDIS), where collinear factorisation is expected to be valid as well~\cite{Collins:1997sr}. Assuming in addition the validity of proton vertex factorisation~\cite{Ingelman:1984ns}, diffractive processes are described by the exchange of collective colourless partonic states, such as the pomeron. Diffractive parton distribution functions (DPDFs) are extracted from diffractive data~\cite{Aktas:2006hy,Chekanov:2009aa}. Similarly to the normal PDFs, the DPDFs are expected to evolve as a function of the scale as predicted by the DGLAP equations. QCD analyses of diffractive data show that gluons constitute the main contribution to the DPDFs~\cite{Aktas:2006hy,Chekanov:2009aa}. To date, analyses of HERA data support the validity of the collinear factorisation theorem in DDIS as evidenced by experimental results on inclusive production~\cite{Aktas:2006hy,Chekanov:2009aa}, dijet production~\cite{Aktas:2007hn,Chekanov:2007aa,Aktas:2007bv,Aaron:2011mp,Chekanov:2009aa,Andreev:2014yra,Andreev:2015cwa} and $\dstar$ production~\cite{Adloff:2001wr,Chekanov2002244,Chekanov20033,Aktas:2006up,Chekanov2007}.

Here, a new measurement of $\dstar(2010)$ meson production in DDIS is presented, where the $\dstar$ is reconstructed in the $\dstar \rightarrow K \pi\pi$ decay channel. The $\dstar$ meson originates from the fragmentation of a charm quark, which is produced at HERA energies mainly via the boson-gluon-fusion ($\gamma^{*}g \rightarrow c\bar{c}$) process. Hence, the gluon content of the pomeron can be accessed directly, and allows the collinear factorisation to be tested. Compared to the previous H1 publication~\cite{Aktas:2006up} the analysis presented corresponds to a sixfold increase in the integrated luminosity. 
\newpage

\section{Kinematics of Diffractive Deep Inelastic Scattering}

\begin{figure}[t]
\center
\epsfig{file=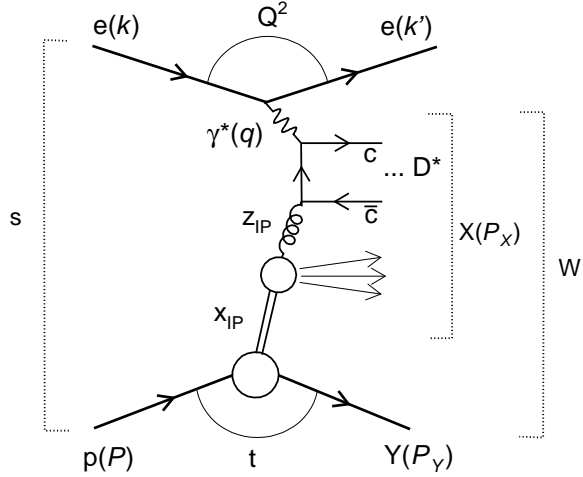 ,width=9cm}
\setlength{\unitlength}{1cm}
\caption{The leading order diagram for open charm production in diffractive DIS at HERA in the picture of collinear and proton vertex factorisation.}
\label{fig:ccbar} 
\end{figure}

The standard DIS kinematics is described in terms of the invariants
\begin{eqnarray}
s = (k+P)^{2} ,\hspace*{0.7cm}
Q^2 = -q^{2}  ,\hspace*{0.7cm}
y = \frac{q\cdot P}{k\cdot P}  ,\hspace*{0.7cm}
W^2 = (q+P)^{2} ,\hspace*{0.7cm}
x = \frac{Q^{2}}{2\,q\cdot P} \ ,
\label{sqyx}
\end{eqnarray}
where the four-vectors are indicated in figure~\ref{fig:ccbar}. Here $s$ is the square of the total centre-of-mass energy of the collision, $Q^{2}$ the photon virtuality, $y$ the scattered electron inelasticity, $W^2$ the centre-of-mass energy squared of the $\gamma^{*}p$ system and $x$ the Bjorken scaling variable.

Given the two hadronic systems $X$ and $Y$, separated by a large rapidity gap, diffractive kinematic variables are defined as follows:
\begin{eqnarray}
M_{X}^2 = (P_{X})^2 ,\hspace*{0.5cm}
M_{Y}^2 = (P_{Y})^2 ,\hspace*{0.5cm}
t = (P-P_{Y})^{2} ,\hspace*{0.5cm}
x_{\pom} = \frac{q\cdot (P-P_{Y})}{q\cdot P}. 
\label{mxmytxpom}
\end{eqnarray}
In inclusive DDIS, where $M_{X}$ and $M_{Y}$ are the invariant masses of the systems $X$ and $Y$, respectively, $t$ is the squared four-momentum transfer at the proton vertex and $x_{\pom}$ the fraction of the proton's longitudinal momentum transferred to the system $X$. In open charm production, the $z_{\pom}$ variable is defined as
\begin{eqnarray}
z_{\pom} = \frac{\hat{s} + Q^{2}} {M_{X}^{2} + Q^{2}}.
\label{eq:zpom}
\end{eqnarray}
It represents, in leading order, the pomeron's momentum fraction participating in the $\gamma^{*}g \rightarrow c\bar{c}$ hard process. The variable $\hat{s}$ denotes the centre-of-mass energy squared of the hard process, corresponding to the centre-of mass energy of the $c\bar{c}$ quark pair in figure~\ref{fig:ccbar}.

\section{Monte Carlo Models and Fixed Order QCD Calculations}
\label{sec:MCandNLO}
The diffractive and non-diffractive processes are modelled with the RAPGAP 
Monte Carlo event generator~\cite{JUNG1995147}. The generated Monte Carlo 
events are subjected to a detailed H1 detector response simulation based on 
GEANT-3~\cite{Brun:1987ma}. These simulated samples are passed through the 
same analysis chain as used for data and are used to correct the data for 
detector effects.

Diffractive events are simulated with leading (pomeron) and sub-leading (reggeon, $\reg$) exchanges based on the H1 2006 DPDF Fit B~\cite{Aktas:2006hy} diffractive parton density parameterisation obtained from a previous QCD analysis of inclusive diffractive data, convoluted with leading order matrix elements for open charm production via photon-gluon fusion. The contribution of non-diffractive processes to open charm production is simulated using RAPGAP in non-diffractive mode with the CTEQ6L PDF set~\cite{Pumplin:2002vw}. Higher order QCD effects are modelled through parton showers in the leading-log approximation. QED radiation effects are simulated with the HERACLES program~\cite{Heracles} interfaced to RAPGAP. To assess the effect of QED radiation a diffractive sample without QED radiation was also generated. Fragmentation is performed using the Lund string model~\cite{Andersson:1983ia} where all decay channels of the charm quark are included. The longitudinal part of the fragmentation function is reweighted according to the Kartvelishvili parameterisation $D(z) \sim z^{\alpha}(1-z)$~\cite{Kartvelishvili} with an appropriate choice of $\alpha$~\cite{Aaron:2008ac}. The simulated events contain both elastic ($ep \rightarrow eXp$) and proton dissociative ($ep \rightarrow eXY$) processes. The two fractions are normalised relative to each other~\cite{Aaron:2010aa},
\begin{equation}
{\sigma(M_{Y} < 1.6~{\rm GeV})} / {\sigma(M_{Y} = m_{p})}  = 1.20 \pm 0.11.
\label{eq:pdnorm}
\end{equation}
Here $\sigma(M_{Y} = m_{p})$ denotes the contribution in which the system $Y$ contains only a leading proton, whereas $\sigma(M_{Y} < 1.6~{\rm GeV})$ also includes contributions from proton dissociation processes integrated up to mass $M_{Y} = 1.6~{\rm GeV}$. The simulated physics events are reweighted in the generated kinematics in order to reach good agreement with data at the reconstructed level as will be shown in section~\ref{sec:eventselection}.

Predictions for $\dstar$ cross sections in next-to-leading-order (NLO) QCD precision are obtained from the HVQDIS~\cite{Harris:1995tu,Harris:1997zq} program adapted for diffraction. The calculation relies on collinear factorisation using H1 2006 DPDF Fit B NLO parton density functions involving gluons and light quarks in the quark singlet (fixed-flavour-number scheme). Massive charm quarks are produced via $\gamma^{*}$-gluon fusion with the QCD scale parameter set to $\Lambda_{5} = 0.228~{\rm GeV}$, which corresponds to a 2-loop $\alpha_{s}(M_{Z})=0.118$, as was used in the DPDF extraction. The charm quarks are fragmented independently into $\dstar$ mesons with $f(c\rightarrow \dstar) = 0.235 \pm 0.007$~\cite{Gladilin:1999pj} in the $\gamma^{*}p$ rest frame using the Kartvelishvili parameterisation with parameters suited for use with HVQDIS~\cite{Aaron:2008ac}. The factorisation and renormalisation scales are set to $\mu_{r} = \mu_{f} = \sqrt{Q^{2} + 4m_{c}^{2}}$ with the value $m_{c} = 1.5{\rm~GeV}$ for the charm pole mass. The uncertainties arising from the choice of scales are estimated by simultaneously varying them by factors of $0.5$ and $2$. The uncertainty introduced in the calculation caused by the uncertainty of $m_c$ is evaluated by varying $m_{c}$ to $1.3~{\rm GeV}$ and $1.7~{\rm GeV}$. The Kartvelishvili parameters are varied within their uncertainties~\cite{Aaron:2008ac}. The DPDF uncertainties are estimated by propagating the eigenvector decomposition of the errors of the DPDF parameterisation. The individual sources of uncertainties are added in quadrature separated for up and down variations of the cross sections. The contribution of $B$-hadrons due to beauty fragmentation to the diffractive $\dstar$ cross section is neglected; it is expected to be less than $3\%$ for non-diffractive DIS (see \cite{Aktas:2006py}) and even smaller for the diffractive production.

The HVQDIS calculation is performed also in the non-diffractive mode using the CT10F3 proton PDF set~\cite{Lai:2010vv}. It is used for comparisons of predictions with measurements of the diffractive to inclusive cross section ratio (section~\ref{sec:resultsdifinclratio}). The calculation is done following the one used for comparison with inclusive $\dstar$ data~\cite{Aaron:2011gp}. The uncertainties from the choice of scales and $m_{c}$ as well as the fragmentation uncertainty are evaluated in the same manner as for the diffractive calculation. The uncertainty of the CT10F3 PDF set is not considered for this analysis but is expected to be small in comparison to the DPDF uncertainties.

\section{Experimental Technique}

\subsection{The H1 Detector}
\label{sec:H1 detector}

A detailed description of the H1 detector can be found elsewhere~\cite{Abt:1996hi}. 
Here, a brief account of the detector components most relevant to the 
present analysis is given. The H1 coordinate system is defined such that 
the origin is at the nominal $ep$ interaction point and the polar angle
$\theta = 0$ and the positive $z$ axis correspond to the direction of the 
outgoing proton beam. The region $\theta < 90^\circ$, which has
positive pseudorapidity $\eta = - \ln \tan\theta /2$, is referred to as the 
`forward' hemisphere.

The {\it ep} interaction point in H1 is surrounded by the central tracking system, which includes silicon strip detectors~\cite{Pitzl:2000wz} as well as two large concentric drift chambers. These chambers cover a region in polar angle $20^{\circ} < \theta < 160^{\circ}$ and provide a resolution of $\sigma(P_{T})/P_{T} = 0.006P_{T}/{\rm~GeV} \oplus 0.02$. They also  
provide triggering information~\cite{Schoning:2006zc,Baird:2001sz}. The forward tracking detector, a set of drift chambers with sense wires oriented perpendicular to the $z$ axis, extends the acceptance of the tracking system down to $7^{\circ}$ in polar angle. The central tracking detectors are surrounded by a finely segmented liquid argon (LAr) sampling calorimeter covering $-1.5 < \eta < 3.4$. Its resolution is $\sigma(E)/E = 0.11/\sqrt{E/{\rm~GeV}} \oplus 0.01$ in its electromagnetic part and $\sigma(E)/E = 0.50/\sqrt{E/{\rm~GeV}} \oplus 0.02$ for hadrons, as measured in test beams~\cite{Andrieu:1993xn,Andrieu:1994yn}.

The central tracker and LAr calorimeter are placed inside 
a large superconducting solenoid, which provides a uniform magnetic 
field of 1.16 T. The backward region $-4 < \eta < -1.4$ is covered by 
a lead-scintillating fiber calorimeter (SpaCal,~\cite{Appuhn:1996na}) with electromagnetic and hadronic sections. In the present analysis the energy and angle of the scattered electron is measured in the electromagnetic section of the SpaCal. It has an energy resolution for electrons $\sigma(E)/E = 0.07 / \sqrt{E/{\rm~GeV}} \oplus 0.01 $ as measured in test beams~\cite{Nicholls:1995di}.

Information from the central tracker and the LAr and SpaCal calorimeters 
is combined in an energy flow reconstruction algorithm which 
yields a list of hadronic final state objects \cite{energyflowalgorithm1,energyflowalgorithm2}.
For these objects a calibration is applied ensuring the relative agreement of hadronic energy scale between the data and simulations at $1 \%$ accuracy~\cite{hfscalibration}. 

In the forward region the H1 detector is equipped with drift chambers
comprising the forward muon detector (FMD, $1.9 < \eta < 3.7$). The forward tagger system (FTS) is a set of scintillators surrounding the beam pipe at several locations along the proton beamline, downstream of the H1 main detector. The FTS station at $28 {\rm~m}$ covering the range $6.0 < \eta < 7.5$ is used in this analysis. Both FMD and FTS are sensitive to the very forward energy flow and improve the selection of large rapidity gap events.

The luminosity is measured via the Bethe-Heitler bremsstrahlung process $ep \rightarrow ep\gamma$, with the final state photon detected by a photon detector located close to the beam pipe at position $z=-103~{\rm m}$. The precision of the integrated luminosity determination is improved in a dedicated analysis of the QED Compton process~\cite{Aaron:2012kn}.

\subsection{Event selection}
\label{sec:eventselection}
The analysis is based on data collected by H1 in the $2005-2006$ electron and the $2006-2007$ positron running periods with $\sqrt{s}=319~{\rm GeV}$, where the proton and lepton beam energies are $920~{\rm GeV}$ and $27.6~{\rm GeV}$, respectively. The corresponding integrated luminosity is $287\text{ pb}^{-1}$. The events are triggered on the basis of a scattered electron signal in the SpaCal calorimeter together with at least one track above the $900$ MeV transverse momentum threshold in the drift chambers of the central tracker.

\subsubsection{Diffractive DIS selection and reconstruction of kinematics}
The momentum transfer $Q^{2}$ and inelasticity $y$ are reconstructed using the electron-$\Sigma$ method~\cite{Adloff:1997sc} which combines information on the scattered electron candidate and hadronic final state (HFS) kinematics. This choice optimises the resolution for these observables. The measurement phase space in $Q^{2}$ and $y$ is chosen to be
\begin{equation}
5 < Q^{2} < 100~{\rm GeV}^{2}\,,\,\,\, 0.02 < y < 0.65.
\label{eq:q2ycuts}
\end{equation}

The selection of diffractive events is based on the presence of a forward large 
rapidity gap (LRG), which is primarily provided by a cut on the pseudorapidity of the most forward cluster in the LAr calorimeter above the $800$ MeV energy threshold,  $\eta_{\mbox{\tiny max}} < 3.2$.

The variable $x_{\pom}$ is reconstructed as
\begin{equation}
x_{\pom} = \frac{M_{X}^{2} + Q^{2}}{W^{2} + Q^{2}},
\label{eq:xpomrec}
\end{equation}
where $W$ is calculated as $W=\sqrt{ys - Q^{2}}$. The invariant mass of the hadronic final state, $M_{X}$, is determined as follows
\begin{equation}
M_{X} = f_{\mbox{\tiny corr}}(\eta_{\mbox{\tiny max}}) \sqrt{(P_{X})^{2}},
\label{eq:mxrec}
\end{equation}
where $P_{X}$ is the reconstructed four-momentum of the hadronic final state and $f_{\mbox{\tiny corr}}$ is an $\eta_{\mbox{ \tiny max}}$ dependent factor introduced in order to correct for detector losses at large $\eta$. It is determined from simulations yielding $16 \%$ enhancement factor on average. The range of the reconstructed $x_{\pom}$ values is limited to $x_{\pom} < 0.03$.

The variable $z_{\pom}$ is reconstructed using in addition the $\dstar$ candidate four-momentum. This variable is denoted $z_{\pom}^{obs}$ and is defined as
\begin{equation}
z_{\pom}^{obs} = \frac{(M_{c\bar{c}}^{2})^{obs} + Q^{2} }  { {M_{X}^{2} + Q^{2}} }\text{ , with  } (M_{c\bar{c}}^{2})^{obs} = \frac{1.2\,p_{\perp \dstar}^{* 2} + m_{c}^{2}} {z(1-z)} \text{  and  } z = \frac{(E-p_{z})_{\dstar}^{\text{(lab)}}}{2yE_{e}}
\label{eq:zpomobs}
\end{equation}
where $(M_{c\bar{c}}^{2})^{obs}$ is an estimate of $\hat{s}$ in equation~\ref{eq:zpom}. $(M_{c\bar{c}}^{2})^{obs}$ is reconstructed from the $\dstar$ kinematics. This is done in close analogy to the $x_g^{\text{obs}}$ measurement in inclusive $\dstar$ production ~\cite{Adloff:1998vb}. The term $1.2\,p_{\perp \dstar}^{* 2}$ is an approximation to the value of the transverse momentum squared of the charm quark in the $\gamma^{*}p$ rest frame. The observable $z$ denotes the inelasticity of the $\dstar$ meson which is calculated in the laboratory frame using the difference of the energy and the longitudinal momentum, $(E-p_{z})_{\dstar}$, of the D* meson. The factor $1.2$ is introduced to ensure $z_{\pom}^{\text{true}} \approx z_{\pom}^{\text{obs}}$ on average, as deduced in studies of generated events using RAPGAP.

The activity in the FTS and the FMD is required not to exceed the noise levels monitored with an event sample triggered independently of detector activity. Noise effects are also propagated into the simulation of the detector response in a similar manner. The diffractive event selection requirements ensure that the analysed sample is dominated by $ep \rightarrow eXp$ processes at small $|t|$ with an intact proton in the final state, often called proton elastic processes. A small admixture of events is present with leading neutrons or low $M_{Y}$ baryon excitations, referred to as proton dissociation contributions (PD). The values of $M_Y$ and $t$ are not reconstructed explicitly. However, as the diffractive selection rejects events at large $M_{Y}$ or large $\left| t \right|$, the measurement is corrected to the $M_{Y} < 1.6 {\rm~GeV}$ and $\left| t \right| < 1 {\rm~GeV}^{2}$ range.



\boldmath
\subsubsection{$\dstar$ selection}
\unboldmath

The detection of $\dstar$ mesons is based on the full reconstruction of its decay products in the `golden channel'
\begin{equation}
D^{*+} \rightarrow D^{0} \pi^{+}_{slow} \rightarrow  (K^{-} \pi^{+}) \pi^{+}_{slow} \quad (+C.C.),
\label{eq:goldenchannel}
\end{equation}
with a branching ratio of $(2.66 \pm 0.03)\%$~\cite{Agashe:2014kda}. Tracks reconstructed in the central tracker are used to identify the decay products. The kaon and pion candidate tracks from $D^{0}$ decays are required to satisfy $p_{t} > 0.3 {\rm~GeV}$ while the slow pion candidate track is required to have $p_{t} > 0.12~{\rm GeV}$, where $p_{t}$ is the transverse momentum of the reconstructed track in the laboratory frame. In order to suppress combinatorial background as well as contributions of other decay channels with at least three charged decay products, called reflections, the invariant mass of the $K^{\mp} \pi^{\pm}$ pair is required to be in agreement with the nominal $D^{0}$ mass ($1864.83 \pm 0.05 {\rm~MeV}$,~\cite{Agashe:2014kda}) within a window of $\pm 80$~MeV. The kinematics of the $\dstar$ meson candidate reconstructed from the $K^{\mp} \pi^{\pm} \pi^{\pm}_{slow}$ system is restricted to the range $p_{t,\dstar} > 1.5 {\rm~GeV}$ and $\left| \eta_{\dstar} \right| < 1.5$.

\begin{figure}[hhh]
\center
\epsfig{file=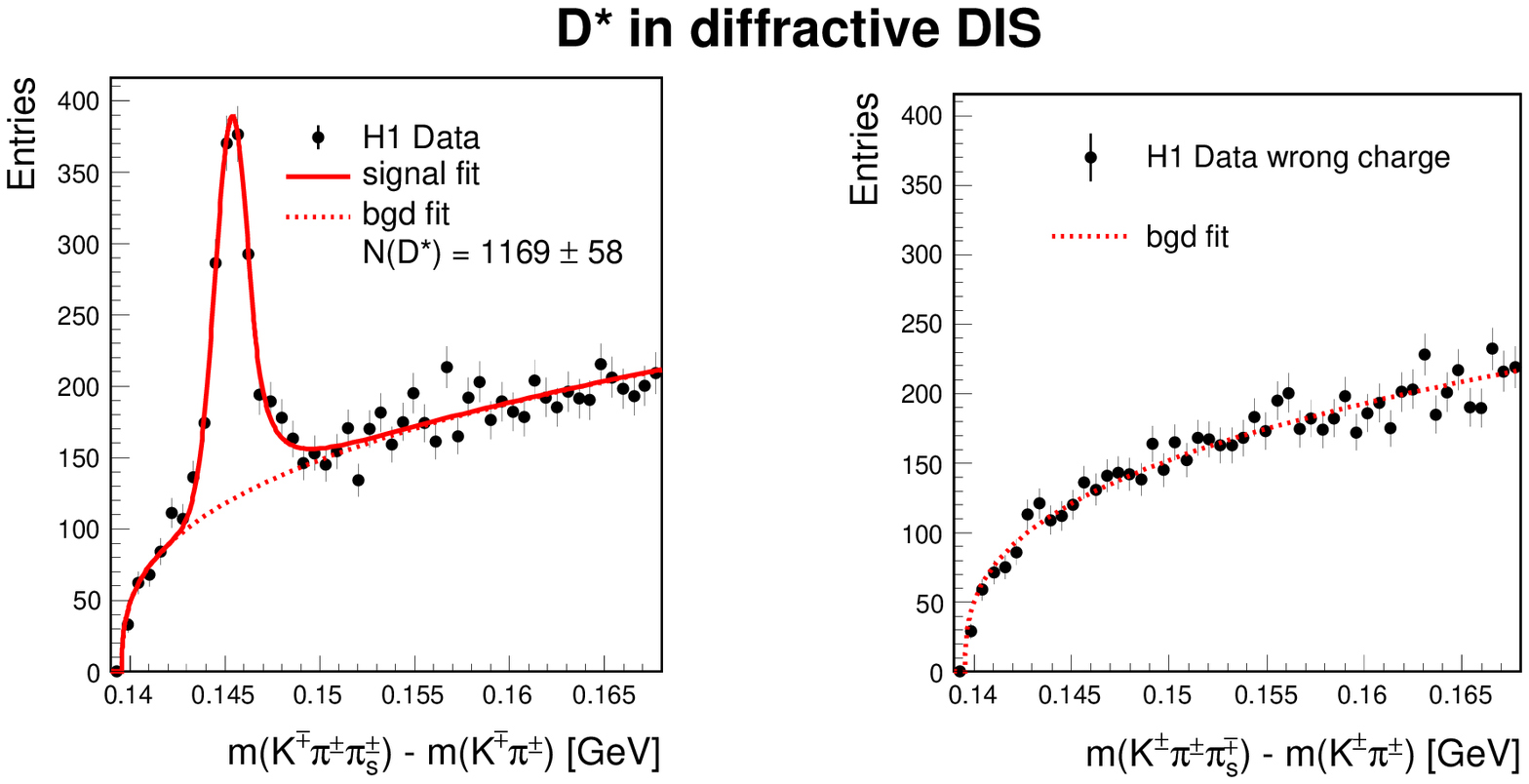 ,width=16cm}
\setlength{\unitlength}{1cm}
\caption{The $\Delta m$ distributions in the data for the right charge sample with the combined signal and background fit indicated by the solid and dotted line, respectively, is shown in the left figure. The wrong charge sample with the background-only fit, performed simultaneously under the assumption of identical background shape in the right charge combinations, is shown in the right figure as the dotted line.}
\label{fig:totaldatafit} 
\end{figure}

The mass difference $\Delta m = m(K^{\mp} \pi^{\pm} \pi^{\pm}_{slow})-m(K^{\mp} \pi^{\pm})$ is used to determine the $\dstar$ signal. It is expected to peak near $\Delta m=0.145~{\rm GeV}$~\cite{Agashe:2014kda}. The wrong charge combinations $K^{\pm} \pi^{\pm} \pi^{\mp}_{slow}$ selected with otherwise unchanged criteria do not contribute to the signal, they are, however, utilised to constrain the shape of the combinatorial background. The right and wrong charge $\Delta m$ distributions are fitted simultaneously by means of an unbinned extended likelihood fit using RooFit~\cite{RooFit} and ROOT~\cite{Brun:1997pa}. The Crystal Ball~\cite{crystalball} and Granet~\cite{Granet:1977px} probability distribution functions are used for modelling the signal and background, respectively. The fit to the total sample of selected $\dstar$ candidates is shown for the right and wrong charge combinations in figure~\ref{fig:totaldatafit}. The fit to the total number of $\dstar$ mesons in the data yields $N(D^{*}) = 1169 \pm 58$. The observed width is dominated by experimental effects.

The fits are repeated in bins of reconstructed kinematic quantities. Figure~\ref{fig:controlplots} shows the $\dstar$ yields determined as a function of the variables $Q^2$, $y$, $\text{log}_{10}(x_{\pom})$, $z_{\pom}^{obs}$, $p_{t,\dstar}$ and $\eta_{\dstar}$. The $N(\dstar)$ distributions are well described by the reweighted simulation. The fraction of proton dissociation processes adjusted globally (as given by equation~\ref{eq:pdnorm}) is largely independent of the kinematics. The reggeon contribution is generally small and reaches $5 \%$ at large $\xpom$. The non-diffractive background contribution is below $1 \%$ level and is not shown.

\subsection{Cross section measurement}
\label{sec:crosssection}
The number of fitted $\dstar$ mesons is corrected for trigger inefficiency, detector effects due to limited acceptance and resolution, the branching ratio of the golden channel, and the contribution of reflections and higher order QED processes at the lepton vertex. The bin averaged $\dstar$ cross section in bin $i$ of a generic variable $x$ in the phase space defined in table~\ref{tab:ddisdstarcuts} is measured as
\begin{equation}
 \left( \frac{ {\rm d} \sigma } { { \rm d} x }  \right) _{i} = \frac{N_{i}^{\rm data} - N_{i}^{\rm sim, bgr} } {\mathcal{L}_{\rm int} \,\, \Delta^{x}_{i} \,\, B_{r} \,\, \varepsilon_{\rm trigg}  \,\, A_{i}} \, C^{\mbox{\tiny QED}}_{\mbox{\tiny corr},i},
\label{eq:binbybinformula}
\end{equation}
where
\begin{itemize}
  \item $N_{i}^{\rm data}$ is the number of $\dstar$ mesons from the fit passing the experimental cuts in the data.

  \item $N_{i}^{\rm sim, bgr}$ is the number of $\dstar$ mesons from the fit to simulated events passing the experimental cuts while being generated outside the phase space  (table~\ref{tab:ddisdstarcuts}) of the measurement.
  
  \item $A_{i}$ is the acceptance correction factor accounting for effects related to the transition from the hadron level to the detector level determined from MC simulations.

  \item $\mathcal{L}_{\rm int}$ is the  integrated luminosity of the data.
  
  \item $B_{r}$ is the branching ratio of the golden decay channel.
  
  \item $\varepsilon_{\rm trigg}$ is the trigger efficiency.

  \item  $C^{\mbox{\tiny QED}}_{\mbox{\tiny corr},i}$ are corrections for QED radiation defined as $\sigma^{\mbox{\tiny QED-off}} / \sigma^{\mbox{\tiny QED-on}}$ as obtained from Monte Carlo generated events, where $\sigma^{\mbox{\tiny QED-off}}$ ($\sigma^{\mbox{\tiny QED-on}}$) is the bin-integrated cross section predicted by RAPGAP with QED radiation turned off (turned on) as described in section~\ref{sec:MCandNLO}.
  
  \item $\Delta_{i}^{x}$ is the bin width of the $i$-th bin of $x$.
\end{itemize}

The acceptance corrections, $A_{i}$, are defined as
\begin{equation}
A_{i} = \frac{ N_{i}^{\rm sim} - N_{i}^{\rm sim, bgr} } {n_{i}^{\rm sim}},
\label{eq:acceptance}
\end{equation}
 where, for a given bin $i$, $N_{i}^{\rm sim}$ is the fitted number of $\dstar$ mesons  passing the experimental cuts in the simulation of MC generated events encompassing all charm quark decay channels as well as all $\dstar$ decay channels, $N_{i}^{\rm sim, bgr}$ is defined above and $n_{i}^{\rm sim}$ is the MC generated number of $\dstar$ mesons decaying solely via the golden channel with the event kinematics inside the phase space defined in table~\ref{tab:ddisdstarcuts}.

\begin{table}[h]
\centering
\begin{tabular}{|c|}
\hline
\multicolumn{1}{|c|}{DIS phase space}     \\
\hline 
$ 5 < Q^{2} <  100~{\rm GeV}^{2}$ \\ 
$0.02 < y < 0.65$ \\
\hline
\multicolumn{1}{|c|}{$D^{*}$ kinematics}  \\ 
\hline 
$p_{t,D^{*}} > 1.5~{\rm GeV}$ \\
$-1.5 < \eta_{D^{*}} < 1.5 $ \\
\hline
\multicolumn{1}{|c|}{Diffractive phase space}     \\
\hline 
$x_{\pom} < 0.03$ \\
$M_{Y} < 1.6~{\rm GeV}$ \\
$|t| < 1~{\rm GeV}^{2} $  \\
\hline
\end{tabular}
\caption{Definition of the phase space of the cross section measurement.}
\label{tab:ddisdstarcuts}
\end{table}

\subsection{Systematic uncertainties}
\begin{figure}[hhh]
\center\epsfig{file=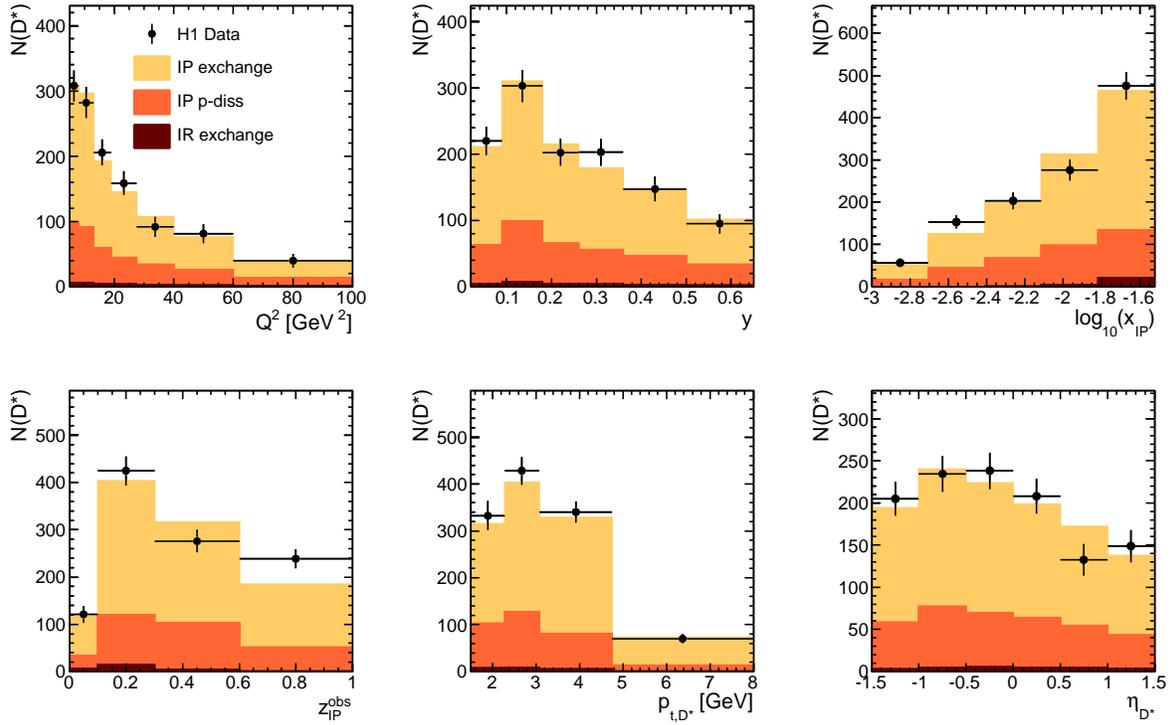 ,width=16cm}
\setlength{\unitlength}{1cm}
\caption{The differential $N(\dstar)$ distributions obtained from $\Delta m$ fits to the data and simulation as a function of $Q^2$, $y$, $\text{log}_{10}(x_{\pom})$, $z_{\pom}^{obs}$, $p_{t,\dstar}$ and $\eta_{\dstar}$. The data are represented with dots. The contributions of individual processes in the simulation, reweighted RAPGAP, are indicated by filled histograms as follows; elastic proton pomeron exchange (light orange), proton dissociation (dark orange) and reggeon exchange (dark red).}
\label{fig:controlplots} 
\end{figure}

Experimental and model uncertainties are propagated to the differential and the integrated $\dstar$ cross section measurements. In the following only the effects on the integrated $\dstar$ cross sections are quantified~\footnote{A detailed analysis of the systematic uncertainties is available \\
\url{http://www-h1.desy.de/publications/H1publication.short_list.html}.}.
\begin{itemize}
  \item The energy scale (polar angle) of the scattered lepton is known to the $1\%$ ($1~{\rm mrad}$) level resulting in a $0.5 \%$ ($1.5 \%$) uncertainty.


  
  \item The relative energy scale of the hadronic final state is known with a precision of $1\%$ resulting in a $0.06\%$ uncertainty.

  \item Changing the function $f_{\mbox{\tiny corr}}$ (equation~\ref{eq:mxrec}) to the constant $1.16$ results in $2.7 \%$ uncertainty.

  \item There is a certain ambiguity in describing the tails of the $\Delta m$ signal distribution. Choosing a modified Crystal Ball function with an extra Gaussian component for the fits to the $\dstar$ signal has $3.8 \%$ effect.

  \item  The normalisation of the proton dissociative contribution (equation~\ref{eq:pdnorm}) introduces an uncertainty of $7.1 \%$.

  \item In a dedicated study~\cite{marketajansovaLRGuncertainty}, using forward proton tagging data, a $10\%$ uncertainty on the large rapidity gap selection inefficiency is determined, which translates to a $2.4 \%$ uncertainty.
  
  \item The shapes of the generated spectra of $Q^{2}$, $y$, $x_{\pom}$, $p_{t,D^{*}}$ and $t$ are varied independently with the help of multiplicative weights of ${\rm e}^{{}^{+0.007 }_{-0.013}\,Q^{2}/{\rm GeV}^{2}}$, $y^{^{+0.9}_{-1.1}}$, $(x_{\pom})^{^{+0.13}_{-0.16}}$, ${\rm e}^{^{+0.06}_{-0.15}{\,}p_{t,\dstar}/{\rm GeV}}$ and ${\rm e}^{^{+0.8}_{-0.9}{\,}t/{\rm GeV}^{2}}$ resulting in variations of the fitted differential distributions compatible with the data control distributions (figure~\ref{fig:controlplots}). The reweighting is an approach to assess the uncertainties on the data correction procedure stemming from the Monte Carlo model. The resulting uncertainties are $0.5\%$, $0.9\%$, $0.4\%$, $3.7\%$ and $1.1\%$, respectively.
\end{itemize}

The following uncertainties affect only the normalisation of the measurement.
\begin{itemize}
  \item The integrated luminosity is known to $2.7 \%$ and the golden channel branching ratio to $1.1 \%$.
  \item The uncertainty on the trigger efficiency ($98 \%$ on average) is covered by a $2\%$ variation.

  \item The impact of the restriction to the $D^{0}$ mass window  in terms of $N(\dstar)$ yield loss caused by the choice of the $80 {\rm ~MeV}$ value is evaluated. A systematic uncertainty of $2 \%$ covers the observed difference between data and simulation.

  \item The reflections contribute about $3 \%$ to the fitted $N(\dstar)$. The branching fractions of $\dstar$ decaying to reflections are not precisely reproduced in the simulation. The integrated cross section increases by about $1.2 \%$ if recent branching ratios of reflections are used~\cite{Agashe:2014kda}.

  \item The track reconstruction efficiency is known with $1\%$ uncertainty resulting in $3\%$ per $\dstar$.

  \item The contribution of non-diffractive processes is suppressed by the diffractive selection to a level of less than $1 \%$. A conservative uncertainty of $1 \%$ is assigned.
\end{itemize}

\par
The contributions of the individual systematic uncertainties are added in quadrature both for the integrated and differential cross section measurements.

\section{Results}
\label{sec:results}
In the first part of this section the measured integrated and differential cross  sections for $\dstar$ production in diffractive deep inelastic scattering are presented. Theoretical predictions based on next-to-leading order QCD calculations are compared with the data. In the second part ratios of diffractive to non-diffractive $\dstar$ production cross sections are extracted and confronted with theoretical predictions as well as with previous results from HERA.

\boldmath
\subsection{Diffractive $\dstar$ production cross sections}
\unboldmath
\label{sec:resultsdif}
The integrated cross section of $\dstar$  production for the phase space region given in table~\ref{tab:ddisdstarcuts} is measured to be 
\begin{equation}
\sigma_{ep \rightarrow \,e Y \!X(\dstar) } = 314  \pm  23 \text{ (stat.) } \pm 35 \text{ (syst.) }  \text{ pb}.
\label{eq:totaldataxsection}
\end{equation}
This can be compared with the theoretical value calculated in next-to-leading order QCD with the HVQDIS code~\cite{Harris:1995tu,Harris:1997zq} adapted to diffraction using H1 2006 DPDF Fit B and Kartvelishvili fragmentation as described in section~\ref{sec:MCandNLO}.
\begin{equation}
\sigma_{ep \rightarrow \,e Y \!X(\dstar) }^{\text{theory}} = 265 \text{ }^{+54}_{-40} \text{ (scale) } \text{ }^{+68}_{-54} \text{ (}m_{c}\text{) } \text{}^{+7.0}_{-8.2} \text{ (frag.) }  \text{ }^{+31}_{-35} \text{ (DPDF) }    \text{ pb.}
\label{eq:totalnloxsection}
\end{equation}
Within its large uncertainties the prediction is compatible with the measured value, which supports collinear factorisation. However, the prediction depends substantially on the choice of the factorisation and renormalisation scale as well as on the value of the charm mass. Similar conclusions were reached in a previous H1 publication~\cite{Aktas:2006up} albeit within larger uncertainties and in a slightly different kinematic domain.

\begin{figure}[hhh]
\center\epsfig{file=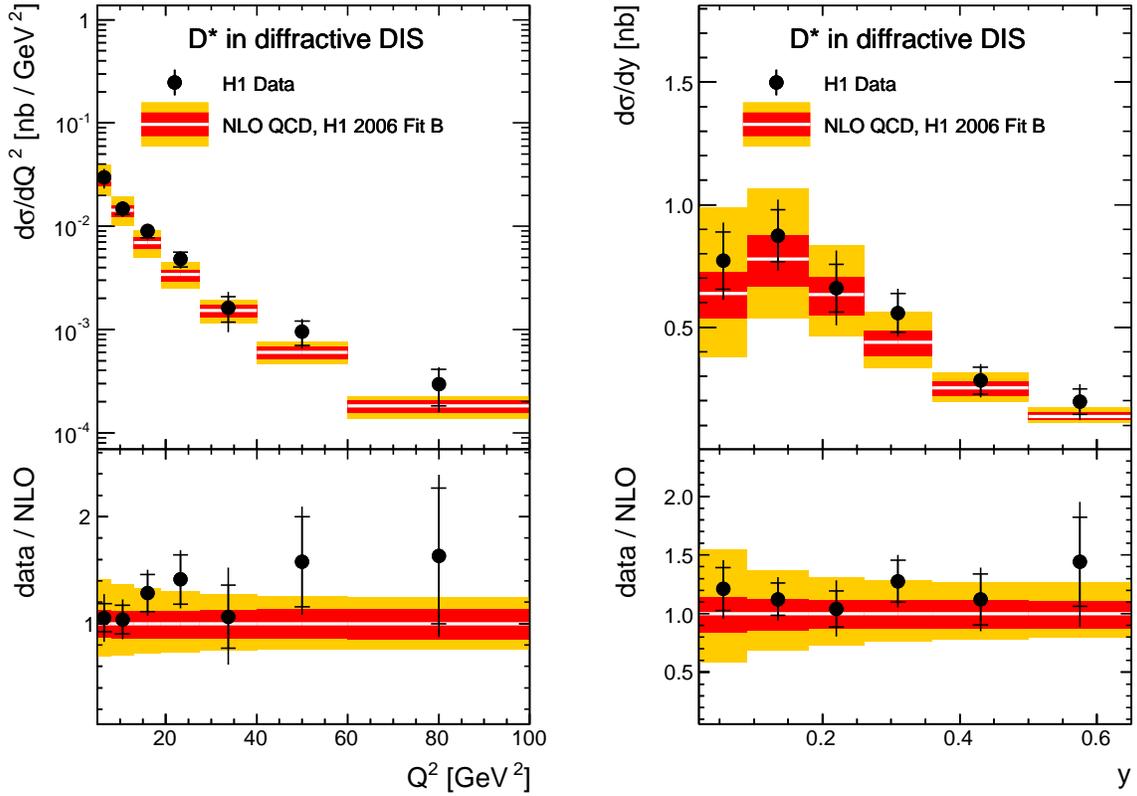 ,width=16cm}
\setlength{\unitlength}{1cm}
\caption{Bin averaged single-differential $\dstar$ cross sections as a function of $Q^2$ and $y$. Data are shown as dots, where the inner error bars indicate statistical uncertainties and the outer error bars represent the statistical and the full set of systematic uncertainties added in quadrature. The central NLO QCD prediction by HVQDIS is shown as a white line inside the coloured bands. The inner band represents the DPDF and fragmentation uncertainties added in quadrature. The outer band represents DPDF, fragmentation, charm mass, factorisation and renormalisation scale uncertainties added in quadrature.}
\label{fig:sigmadis} 
\end{figure}

\begin{figure}[hhh]
\center
\epsfig{file=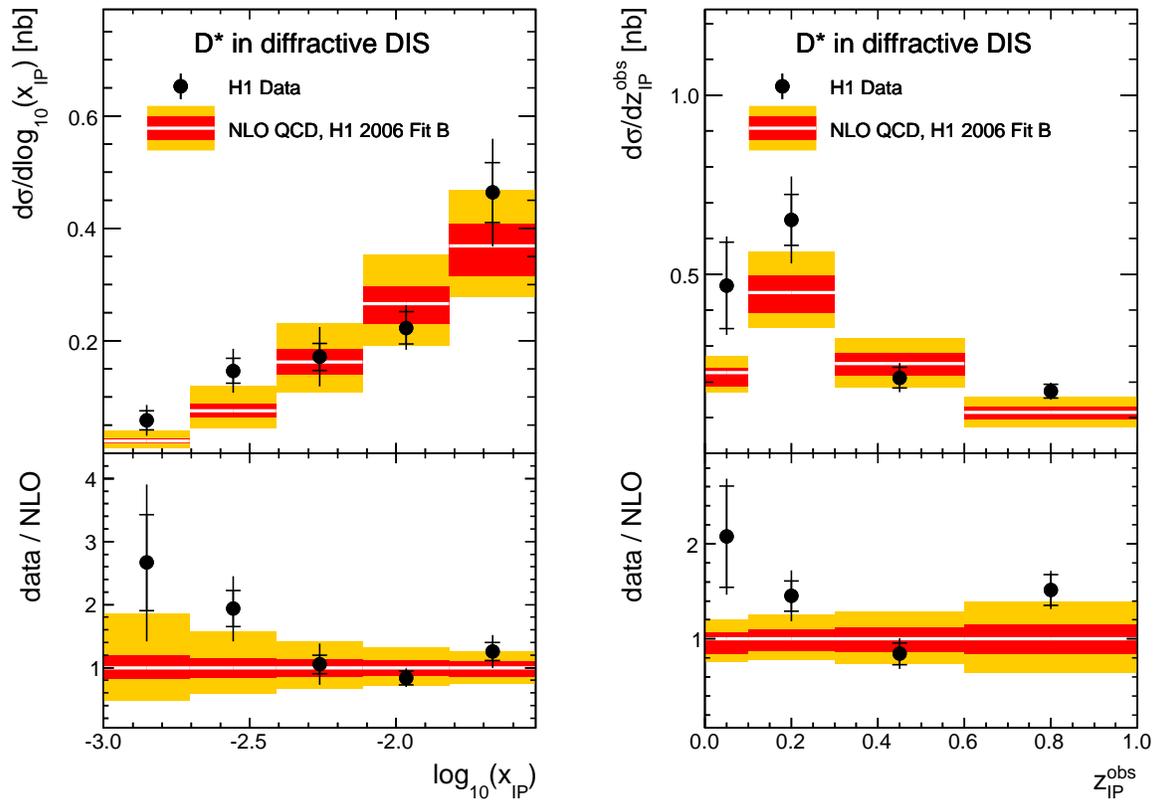 ,width=16cm}
\setlength{\unitlength}{1cm}
\caption{Bin averaged single-differential $\dstar$ cross sections as a function of the diffractive variables $\text{log}_{10}(x_{\pom})$ and $z_{\pom}^{obs}$. Further details are indicated in the caption of figure~\ref{fig:sigmadis}.}
\label{fig:sigmadif} 
\end{figure}

\begin{figure}[hhh]
\center
\epsfig{file=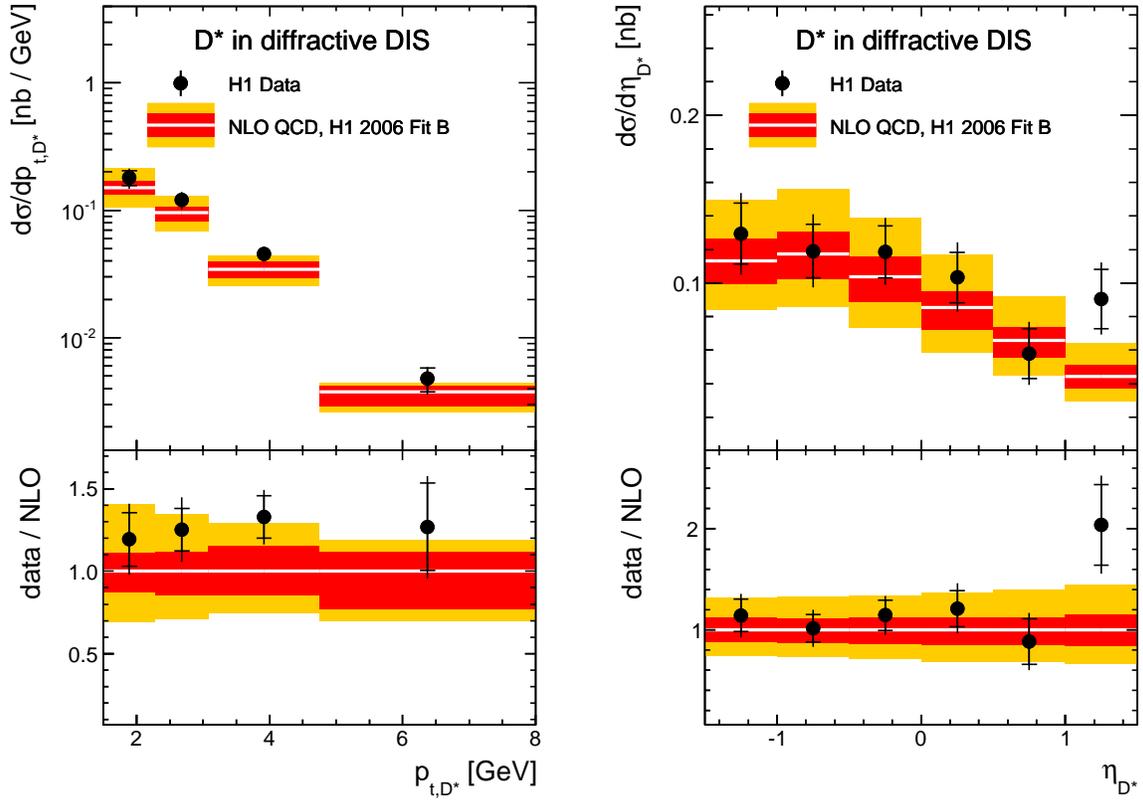 ,width=16cm}
\setlength{\unitlength}{1cm}
\caption{Bin averaged single-differential $\dstar$ cross sections as a function of $p_{t,\dstar}$ and $\eta_{\dstar}$. Further details are indicated in the caption of figure~\ref{fig:sigmadis}.}
\label{fig:sigmadstar} 
\end{figure}

The measured bin averaged single-differential cross sections as a function of $y$, $Q^{2}$, $\text{log}_{10}(x_{\pom})$, $z_{\pom}^{obs}$, $p_{t,\dstar}$ and $\eta_{\dstar}$ are given in table~\ref{tab:diffsigma} and are shown in figures~\ref{fig:sigmadis},~\ref{fig:sigmadif} and ~\ref{fig:sigmadstar} together with the NLO predictions. In order to compare the shapes between data and theory the ratios of data to NLO calculations are also shown.

Figure~\ref{fig:sigmadis} shows that the shape of the measured ${\rm d}\sigma/{\rm d}y$ is well described by the theory. The measured ${\rm d}\sigma/{\rm d}Q^{2}$ might indicate a slightly harder dependence in the data, however, within the large uncertainties the shape is in agreement with the theory. The shape of the ${\rm d}\sigma/{\rm dlog}_{10}(x_{\pom})$ shown in figure~\ref{fig:sigmadif} is satisfactorily described by the prediction given the large relative uncertainties at low $x_{\pom}$ values. The shape of ${\rm d}\sigma/{\rm d}z_{\pom}^{obs}$ shown in figure~\ref{fig:sigmadif} is not described as well by the prediction, however the experimental uncertainties at low $z_{\pom}^{obs}$ are sizeable.
The shapes of ${\rm d}\sigma/{\rm d} p_{t,\dstar}$ and ${\rm d}\sigma/{\rm d}\eta_{\dstar}$ are well described by the theory (see figure~\ref{fig:sigmadstar}). For $\eta_{\dstar} > 1$, however, the theory predicts a value which underestimates the data by about $50 \%$ with a large uncertainty. There is an indication of a similar effect in the corresponding non-diffractive $\dstar$ cross section measurement~\cite{Aaron:2011gp}.

The differential comparison profits from the substantial increase of statistics in the present analysis as compared with previous measurements at HERA.

\subsection{Diffractive fractions}
\label{sec:resultsdifinclratio}
The $\dstar$ and DIS selection criteria given in table~\ref{tab:ddisdstarcuts} are close to those used in the corresponding non-diffractive analysis~\cite{Aaron:2011gp}. The non-diffractive $\dstar$ differential cross sections thus can be used to calculate the diffractive fraction, $R_{D} = {\sigma^{\text{diff}}_{\text{\dstar}}} / {{\sigma^{\text{non-diff}}_{\text{\dstar}}}}$, in the phase space defined in table~\ref{tab:ddisdstarcuts}.

The non-diffractive cross sections~\cite{Aaron:2011gp}, originally given for $0.02 < y < 0.7$, are interpolated to $0.02 < y < 0.65$ using small corrections calculated with HVQDIS. The correction factors reduce the non-diffractive cross sections by about $1.5-3.5 \%$ differentially in $Q^2$, $p_{t,\dstar}$ and $\eta_{\dstar}$ with typical uncertainties of $0.2 \%$. The uncertainties of both the diffractive and non-diffractive cross sections are accounted for in the $R_{D}$ measurement. Integrated over the whole phase space the results are $R_{D} = 6.6 \pm 0.5 {\rm (stat)} \text{ }^{+0.9}_{-0.8} {\rm (syst)}\, \%$ for the data and $R_{D}^{\rm\, theory} = 6.0 {}^{+1.0}_{-0.7} {\rm (scale)} {}^{+0.5}_{-0.4} {\rm (}m_c{\rm )} {}^{+0.7}_{-0.8} {\rm (DPDF)} {}^{+0.02}_{-0.04} {\rm (frag)} \,\%$ for the theoretical prediction. The uncertainties of the theoretical predictions are obtained from simultaneous variations of $m_c$, fragmentation parameters and the factorisation and renormalisation scales. The DPDF uncertainty is also propagated to the prediction.
 
\begin{figure}[hhh]
\center
\epsfig{file=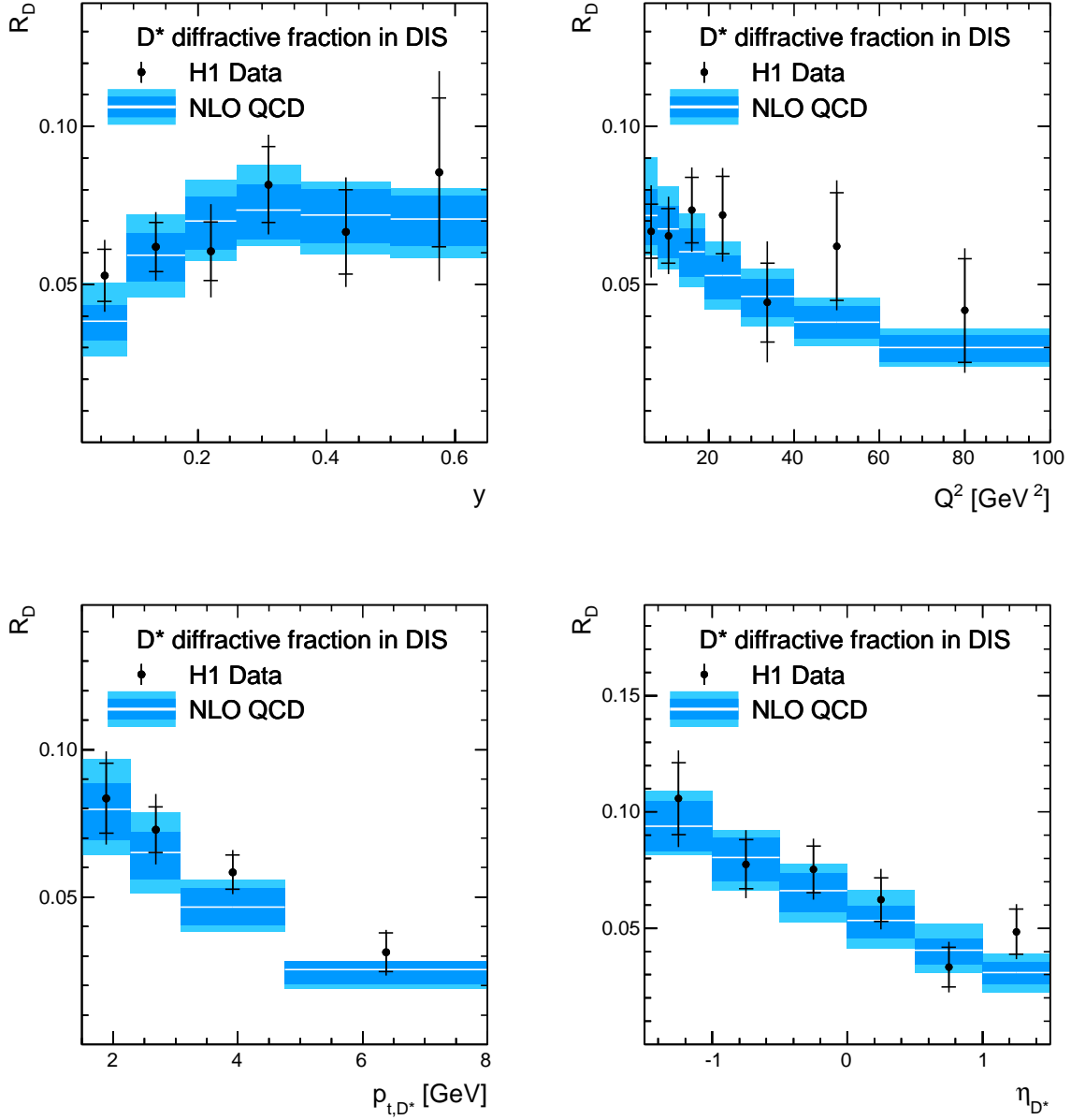 ,width=16cm}
\setlength{\unitlength}{1cm}
\caption{The diffractive fraction, $R_{D}$, measured as a ratio of bin averaged diffractive to non-diffractive $\dstar$ production single differential cross sections in deep inelastic scattering as a function of $y$, $Q^2$, $p_{t,\dstar}$ and $\eta_{\dstar}$. The data ratios are represented with dots, where the inner error bars indicate statistical uncertainties and the outer error bars represent the statistical and systematic uncertainties added in quadrature. The central NLO QCD prediction of $R_D$ by HVQDIS is shown as a white line inside the coloured bands. The inner band represents DPDF uncertainty. The outer band represents effect of the DPDF uncertainty and simultaneous variations of scales, charm mass and fragmentation settings in the diffractive and non-diffractive calculations added in quadrature.}
\label{fig:ratios} 
\end{figure}

The differential fractions $R_{D}(y)$, $R_{D}(Q^2)$, $R_{D}(p_{t,\dstar})$ and $R_{D}(\eta_{\dstar})$ are listed in table~\ref{tab:rdiff} and are shown in figure~\ref{fig:ratios}. Within uncertainties the data do not provide strong evidence for kinematic dependencies of $R_D$ on $Q^2$ or $y$, while at the same time they are also consistent with the kinematic dependencies predicted by theory. The diffractive fraction decreases from $8 \%$ to $3 \%$ with $p_{t,\dstar}$ increasing. The measured dependence of the diffractive fraction on $\eta_{\dstar}$ decreases from $10 \%$ to about $5 \%$ for the highest $\eta_{\dstar}$ values. These shapes are well reproduced by the NLO QCD predictions within the uncertainties. The shapes can be qualitatively understood as follows. Due to the high energy of the leading proton in diffraction ($x_{\pom} < 0.03$) the system $X$ is produced with low masses $M_X$. Less energy is available from the proton side to produce the hard system containing the $\dstar$ meson as compared to the non-diffractive case. Similarly, the fraction is suppressed for small y, i.e. for small energy of the exchanged virtual photon. The $Q^2$ dependence of $R_D$ can be explained by the fact that at high $Q^2$ (higher $x$) the diffractive cross section is suppressed due to a limited $\xpom$ range. Likewise, due to the lack of energy, the events with higher $p_{t,\dstar}$ can be expected to be suppressed in diffraction. The diffractive fraction as a function of $\eta_{\dstar}$ indicates that in diffraction the hard system tends to be produced backwards, due to the kinematics constrained by the presence of a large rapidity gap, or equivalently the $x_{\pom} < 0.03$ condition.

\begin{figure}[hhh]
\center
\epsfig{file=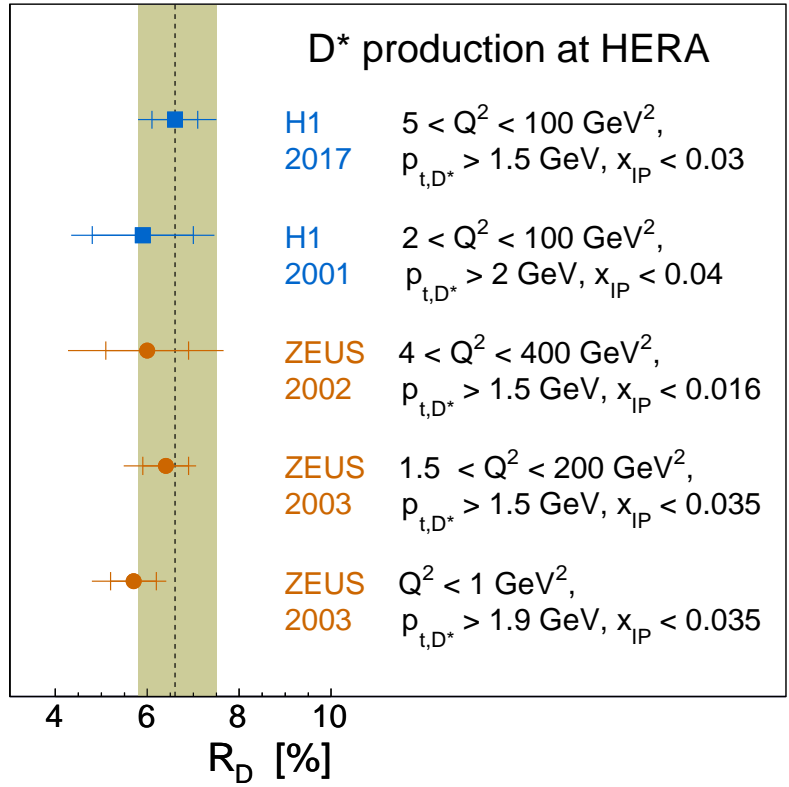 ,width=12cm}
\setlength{\unitlength}{1cm}
\caption{Integrated diffractive fractions measured in $\dstar$ production in the deep-inelastic and the photoproduction ($Q^{2} < 1 {\rm~GeV}^2$) regime as measured at HERA previously \cite{Adloff:2001wr,Chekanov2002244,Chekanov20033,Chekanov2007} and in the present analysis. The inner error bars represent statistical uncertainties, the outer ones the statistical and systematic uncertainties added in quadrature. The dashed line and the shaded band indicate the central value and the total experimental uncertainty of $R_{D}$ of the measurement presented here, respectively.}
\label{fig:totalratios} 
\end{figure}

In figure~\ref{fig:totalratios} the diffractive fraction, integrated over the full phase space, is compared with previous measurements performed at HERA both in the DIS regime~\cite{Adloff:2001wr,Chekanov2002244,Chekanov20033} and in photoproduction~\cite{Chekanov2007}. The average value of $R_{D}$ measured in this article is in agreement with the previous results and within the sizeable experimental uncertainties is observed to be largely independent of the varying phase space constraints in $x_{\pom}$, $Q^2$ and $p_{t,\dstar}$. In particular, the ratios observed in DIS and in photoproduction are compatible with each other.

\section{Conclusions}
\label{sec:conclusions}
Integrated and differential cross sections of $\dstar(2010)$ production in diffractive deep inelastic scattering are measured. The analysis is based on a data sample taken by the H1 experiment at the HERA collider corresponding to an integrated luminosity of $287\text{ pb}^{-1}$. The measured cross sections are compared with theoretical predictions in next to leading order QCD. Compared to the previous measurement in a similar kinematic domain the precision is improved by a factor of two. The new measurements are well described by the predictions within the large theoretical uncertainties which are dominated by variations of scales and the charm quark mass. This supports the validity of collinear factorisation in diffraction.

Measurements of diffractive fractions of $\dstar$ production cross section in deep inelastic scattering are also presented, using non-diffractive cross sections published earlier by H1. The fractions are in agreement with theoretical predictions in next-to-leading order QCD. Although the value of the diffractive fraction is found to decrease at high $p_{t,\dstar}$ and at high $\eta_{\dstar}$ due to limitations of the diffractive phase space, it is observed to be largely independent of other details of the phase space definition. This is confirmed by comparisons to previous measurements of the diffractive fraction.

\clearpage

\section*{Acknowledgements}
We are grateful to the HERA machine group whose outstanding efforts have made this experiment possible. We thank the engineers and technicians for their work in constructing and maintaining the H1 detector, our funding agencies for financial support, the DESY technical staff for continual assistance and the DESY directorate for support and for the hospitality which they extend to the non-DESY members of the collaboration. We would like to give credit to all partners contributing to the EGI computing infrastructure for their support for the H1 Collaboration.
 
We express our thanks to all those involved in securing not only the H1 data but also the software and working environment for long term use allowing the unique H1 dataset to continue to be explored in the coming years. The transfer from experiment specific to central resources with long term support, including both storage and batch systems has also been crucial to this enterprise. We therefore also acknowledge the role played by DESY-IT and all people involved during this transition and their future role in the years to come.


\bibliography{desy17-043}
{}
\bibliographystyle{desy17-043}

\clearpage

\begin{table}[h]
\centering
\begin{tabular}{|ccc|c|c|c|c|}
\hline
\multicolumn{3}{|c|} {$y$} & ${\rm d}\sigma/{\rm d}y\,\,[{\rm pb}]$ & $\delta_{{\rm stat}}\,\,[{\rm pb}]$ & $\delta_{{\rm syst}}\,\,[{\rm pb}]$ & $\delta_{{\rm tot}}\,\,[{\rm pb}]$ \\
\hline 
$0.02$&$\div$&$0.09$ & $770$ & $120$ & $110$ & $160$ \\ 
$0.09$&$\div$&$0.18$ & $870$ & $110$ & $100$ & $150$ \\ 
$0.18$&$\div$&$0.26$ & $660$ & $98$ & $117$ & $152$ \\ 
$0.26$&$\div$&$0.36$ & $558$ & $78$ & $58$ & $97$ \\ 
$0.36$&$\div$&$0.50$ & $282$ & $55$ & $41$ & $68$ \\ 
$0.50$&$\div$&$0.65$ & $197$ & $52$ & $51$ & $73$ \\ 
\hline
\hline
\multicolumn{3}{|c|} {$Q^{2}\,[{\rm GeV}^{2}]$} & ${\rm d}\sigma/{\rm d}Q^{2}\,\,[{\rm pb} / {\rm GeV}^{2}]$ & $\delta_{{\rm stat}}\,\,[{\rm pb} / {\rm GeV}^{2}]$ & $\delta_{{\rm syst}}\,\,[{\rm pb} / {\rm GeV}^{2}]$ & $\delta_{{\rm tot}}\,\,[{\rm pb} / {\rm GeV}^{2}]$ \\
\hline 
$5$&$\div$&$8$ & $29.6$ & $3.7$ & $5.0$ & $6.2$ \\ 
$8$&$\div$&$13$ & $14.8$ & $1.9$ & $1.9$ & $2.7$ \\ 
$13$&$\div$&$19$ & $9.0$ & $1.2$ & $0.9$ & $1.5$ \\ 
$19.0$&$\div$&$27.5$ & $4.81$ & $0.79$ & $0.48$ & $0.92$ \\ 
$27.5$&$\div$&$40.0$ & $1.63$ & $0.45$ & $0.52$ & $0.69$ \\ 
$40$&$\div$&$60$ & $0.95$ & $0.25$ & $0.17$ & $0.30$ \\ 
$60$&$\div$&$100$ & $0.30$ & $0.11$ & $0.07$ & $0.14$ \\ 
\hline
\hline
\multicolumn{3}{|c|}{${\rm log}_{10}(x_{\pom})$} & ${\rm d}\sigma/{\rm d}{\rm log}_{10}(x_{\pom})\,\,[{\rm pb}]$ & $\delta_{{\rm stat}}\,\,[{\rm pb}]$ & $\delta_{{\rm syst}}\,\,[{\rm pb}]$ & $\delta_{{\rm tot}}\,\,[{\rm pb}]$ \\
\hline 
$-3.00$&$\div$&$-2.70$ & $59$ & $17$ & $22$ & $27$ \\ 
$-2.70$&$\div$&$-2.41$ & $147$ & $22$ & $32$ & $39$ \\ 
$-2.41$&$\div$&$-2.11$ & $172$ & $24$ & $47$ & $53$ \\ 
$-2.11$&$\div$&$-1.82$ & $223$ & $29$ & $27$ & $40$ \\ 
$-1.82$&$\div$&$-1.52$ & $464$ & $53$ & $79$ & $96$ \\ 
\hline
\hline
\multicolumn{3}{|c|}{$z_{\pom}$} & ${\rm d}\sigma/{\rm d}z_{\pom}\,\,[{\rm pb}]$ & $\delta_{{\rm stat}}\,\,[{\rm pb}]$ & $\delta_{{\rm syst}}\,\,[{\rm pb}]$ & $\delta_{{\rm tot}}\,\,[{\rm pb}]$ \\
\hline 
$0.0$&$\div$&$0.1$ & $470$ & $120$ & $70$ & $140$ \\ 
$0.1$&$\div$&$0.3$ & $652$ & $71$ & $98$ & $121$ \\ 
$0.3$&$\div$&$0.6$ & $211$ & $29$ & $28$ & $40$ \\ 
$0.6$&$\div$&$1.0$ & $174$ & $19$ & $13$ & $23$ \\ 
\hline
\hline
\multicolumn{3}{|c|}{$p_{t,\dstar}\,[{\rm GeV}]$} & ${\rm d}\sigma/{\rm d}p_{t,\dstar}\,\,[{\rm pb} / {\rm GeV}]$ & $\delta_{{\rm stat}}\,\,[{\rm pb} / {\rm GeV}]$ & $\delta_{{\rm syst}}\,\,[{\rm pb} / {\rm GeV}]$ & $\delta_{{\rm tot}}\,\,[{\rm pb} / {\rm GeV}]$ \\
\hline 
$1.50$&$\div$&$2.28$ & $180$ & $24$ & $22$ & $33$ \\ 
$2.28$&$\div$&$3.08$ & $120$ & $12$ & $14$ & $19$ \\ 
$3.08$&$\div$&$4.75$ & $45.6$ & $4.4$ & $3.5$ & $5.7$ \\ 
$4.75$&$\div$&$8.00$ & $4.8$ & $1.0$ & $0.6$ & $1.2$ \\ 
\hline
\hline
\multicolumn{3}{|c|}{$\eta_{\dstar}$} & ${\rm d}\sigma/{\rm d}\eta_{\dstar}\,\,[{\rm pb}]$ & $\delta_{{\rm stat}}\,\,[{\rm pb}]$ & $\delta_{{\rm syst}}\,\,[{\rm pb}]$ & $\delta_{{\rm tot}}\,\,[{\rm pb}]$ \\
\hline 
$-1.5$&$\div$&$-1.0$ & $129$ & $18$ & $16$ & $24$ \\ 
$-1.0$&$\div$&$-0.5$ & $119$ & $16$ & $15$ & $22$ \\ 
$-0.5$&$\div$&$0.0$ & $119$ & $15$ & $12$ & $20$ \\ 
$0.0$&$\div$&$0.5$ & $103$ & $15$ & $14$ & $21$ \\ 
$0.5$&$\div$&$1.0$ & $58$ & $15$ & $11$ & $19$ \\ 
$1.0$&$\div$&$1.5$ & $91$ & $18$ & $12$ & $22$ \\ 
\hline
\end{tabular}
\caption{Bin averaged hadron level $\dstar$ production cross sections in diffractive DIS as a function of $y$, $Q^{2}, {\rm log}_{10}(x_{\pom})$, $z_{\pom}^{{\rm obs}}$, $p_{t,\dstar}$ and $\eta_{\dstar}$ together with statistical ($\delta_{\rm stat}$), systematic ($\delta_{\rm syst}$) and total ($\delta_{\rm tot}$) uncertainties. The total uncertainties are obtained as the statistical and systematic uncertainties added in quadrature.}
\label{tab:diffsigma}
\end{table}

\begin{table}[h]
\centering
\begin{tabular}{|ccc|c|c|c|c|}
\hline
\multicolumn{3}{|c|}{$y$} & $R_{D}\,\,[\%]$ & $\delta_{\text{stat}}\,\,[\%]$ & $\delta_{\text{syst}}\,\,[\%]$ & $\delta_{\text{tot}}\,\,[\%]$ \\
\hline 
$0.02$&$\div$&$0.09$ & $5.3$ & $0.8$ & $0.8$ & $1.1$ \\ 
$0.09$&$\div$&$0.18$ & $6.2$ & $0.8$ & $0.8$ & $1.0$ \\ 
$0.18$&$\div$&$0.26$ & $6.0$ & $0.9$ & $1.2$ & $1.5$ \\ 
$0.26$&$\div$&$0.36$ & $8.2$ & $1.2$ & $1.0$ & $1.6$ \\ 
$0.36$&$\div$&$0.50$ & $6.7$ & $1.3$ & $1.1$ & $1.7$ \\ 
$0.50$&$\div$&$0.65$ & $8.5$ & $2.4$ & $2.3$ & $3.3$ \\ 
\hline
\hline
\multicolumn{3}{|c|}{$Q^{2}\,[\text{GeV}^{2}]$} & $R_{D}\,\,[\%]$ & $\delta_{\text{stat}}\,\,[\%]$ & $\delta_{\text{syst}}\,\,[\%]$ & $\delta_{\text{tot}}\,\,[\%]$ \\
\hline 
$5$&$\div$&$8$ & $6.7$ & $0.9$ & $1.2$ & $1.5$ \\ 
$8$&$\div$&$13$ & $6.5$ & $0.9$ & $0.9$ & $1.2$ \\ 
$13$&$\div$&$19$ & $7.4$ & $1.0$ & $0.9$ & $1.3$ \\ 
$19.0$&$\div$&$27.5$ & $7.2$ & $1.2$ & $0.8$ & $1.5$ \\ 
$27.5$&$\div$&$40$ & $4.4$ & $1.2$ & $1.5$ & $1.9$ \\ 
$40$&$\div$&$60$ & $6.2$ & $1.7$ & $1.2$ & $2.1$ \\ 
$60$&$\div$&$100$ & $4.2$ & $1.6$ & $1.1$ & $2.0$ \\ 
\hline
\hline
\multicolumn{3}{|c|}{$p_{t,\dstar}\,[\text{GeV}]$} & $R_{D}\,\,[\%]$ & $\delta_{\text{stat}}\,\,[\%]$ & $\delta_{\text{syst}}\,\,[\%]$ & $\delta_{\text{tot}}\,\,[\%]$ \\
\hline 
$1.5$&$\div$&$2.28$ & $8.4$ & $1.2$ & $1.1$ & $1.6$ \\ 
$2.28$&$\div$&$3.08$ & $7.3$ & $0.8$ & $0.9$ & $1.2$ \\ 
$3.08$&$\div$&$4.75$ & $5.8$ & $0.6$ & $0.5$ & $0.8$ \\ 
$4.75$&$\div$&$8.00$ & $3.1$ & $0.7$ & $0.4$ & $0.8$ \\ 
\hline
\hline
\multicolumn{3}{|c|}{$\eta_{\dstar}$} & $R_{D}\,\,[\%]$ & $\delta_{\text{stat}}\,\,[\%]$ & $\delta_{\text{syst}}\,\,[\%]$ & $\delta_{\text{tot}}\,\,[\%]$ \\
\hline 
$-1.5$&$\div$&$-1$ & $10.6$ & $1.5$ & $1.4$ & $2.1$ \\ 
$-1.0$&$\div$&$-0.5$ & $7.8$ & $1.1$ & $1.0$ & $1.5$ \\ 
$-0.5$&$\div$&$0.0$ & $7.5$ & $1.0$ & $0.8$ & $1.3$ \\ 
$0.0$&$\div$&$0.5$ & $6.2$ & $0.9$ & $0.9$ & $1.3$ \\ 
$0.5$&$\div$&$1.0$ & $3.3$ & $0.9$ & $0.7$ & $1.1$ \\ 
$1.0$&$\div$&$1.5$ & $4.9$ & $1.0$ & $0.7$ & $1.2$ \\ 
\hline
\end{tabular}
\caption{The values of diffractive fraction for $\dstar$ production cross sections together with statistical ($\delta_{\rm stat}$), systematic ($\delta_{\rm syst}$) and total uncertainties ($\delta_{\rm tot}$) as a function of $y$, $Q^{2}$, $p_{t,\dstar}$ and $\eta_{\dstar}$. The total uncertainties are obtained as the statistical and systematic uncertainties added in quadrature.}
\label{tab:rdiff}
\end{table}


\end{document}